\documentclass[aps,prl,twocolumn,groupedaddress]{revtex4-1}
\usepackage{amsmath}
\usepackage{graphicx}
\DeclareGraphicsExtensions{.pdf,.eps,.png,.jpg,.mps}
\usepackage{epstopdf}
\usepackage{threeparttable}
\usepackage{color}
\begin{document}
\title{Cleaning Interfaces in Layered Materials Heterostructures}
\author{D. G. Purdie$^1$, N. M. Pugno$^{2,3,4}$, T. Taniguchi$^5$, K. Watanabe$^{5}$, A. C. Ferrari$^{1}$, A. Lombardo$^1$}
\email[]{al515@cam.ac.uk}
\affiliation{$^1$ Cambridge Graphene Centre, University of Cambridge, Cambridge CB3 0FA, UK}
\affiliation{$^2$ Laboratory of Graphene and Bio-inspired Nanomaterials, Department of Civil, Environmental and Mechanical Engineering, University of Trento, Italy}
\affiliation{$^3$ School of Engineering and Materials Science, Queen Mary University of London, UK}
\affiliation{$^4$ Ket-lab, E. Amaldi Foundation, Italian Space Agency, Via del Politecnico, Rome 00133, Italy}
\affiliation{$^5$ National Institute for Materials Science, 1-1 Namiki, Tsukuba 305-0044, Japan}
\begin{abstract}
Heterostructures formed by stacking layered materials require atomically clean interfaces. However, contaminants are usually trapped between the layers, aggregating into blisters. We report a process to remove such blisters, resulting in clean interfaces. We fabricate blister-free regions of graphene encapsulated in hexagonal boron nitride of$\sim5000\mu $m$^{2}$, limited only by the size of the exfoliated flakes. These have mobilities up to$\sim$180000cm$^2$V$^{-1}$s$^{-1}$ at room temperature, and$\sim1.8\times10^6$cm$^2$V$^{-1}$s$^{-1}$ at 9K. We further demonstrate the effectiveness of our approach by cleaning heterostructures assembled using graphene intentionally exposed to polymers and solvents. After cleaning, these samples reach similar high mobilities. We also showcase the general applicability of our approach to layered materials by cleaning blisters in other heterostructures based on MoS$_{2}$. This demonstrates that exposure of graphene to processing-related contaminants is compatible with the realization of high mobility samples, paving the way to the development of fab-based processes for the integration of layered materials in (opto)-electronic devices.
\end{abstract}
\maketitle
\section{Introduction}
The ability to stack single layer graphene (SLG) and other layered materials (LMs) into heterostructures (LMH) has made it possible to create materials where the properties can be tailored by design\cite{PonoNATP2011,BritSCI2012,GorbNATP2012,LeeNATN2014,FerrNAN2014,BonaMT2012,GeorNATN2012}. A number of challenges remain before LMHs can be widely applied, such as the need to use LMs prepared by scalable techniques, like chemical vapor deposition (CVD)\cite{LiSCI2009,BaeNATN2010,YuNatM2011,HaoSCI2013,PetrNL2012}, and to achieve clean interfaces over the entire dimensions of the LMH. One of the most widely studied LMHs is SLG encapsulated in hexagonal boron nitride (hBN)\cite{DeanNATN2010,MayoNL2011,BansNL2016,BansSCIA2015,MayoNL2011,DaubAPL2015,PizzNATC2016,WangSCI2013,DeanNAT2014}. hBN encapsulated SLG can achieve room temperature (RT) mobilities ($\mu$) up to$\sim$150000cm$^2 V^{-1} \ s^{-1}$\cite{WangSCI2013}, over an order of magnitude higher than SLG on SiO$_{2}$\cite{DeanNATN2010}. Furthermore, encapsulation isolates SLG from sources of contamination, such as lithographic polymers and solvents used during device processing\cite{WangSCI2013}, or ambient air\cite{MayoNL2011}, which can otherwise degrade $\mu$\cite{BansSCIA2015,LinNL2012} and increase doping\cite{MayoNL2011}. Thus, hBN encapsulated SLG could enable state of the art performance for a range of applications in high-frequency electronics\cite{MeriIEEE2013,PetrACSN2015,KimACSN2012,FerrNAN2014}, and (opto)electronics\cite{WoesNM2014,GaoNL2015,ShiueNL2015}.

Encapsulated SLG and other LMHs are assembled by first producing the individual LMs on separate substrates, typically Si+SiO$_2$\cite{WangSCI2013}, or polymers, such as Polymethyl methacrylate (PMMA)\cite{KretNL2014,DeanNATN2010}, followed by transfer and stacking to achieve the desired LMHs\cite{WangSCI2013,KretNL2014,DeanNATN2010}. During stacking, contaminants such as hydrocarbons\cite{HaigNATM2012}, air\cite{PizzNATC2016} or water\cite{BampJPC2016,GhorNATC2017}, can become trapped between the layers, aggregating into spatially localized pockets with lateral sizes which vary from tens nm\cite{KhesNATC2016} up to a few $\mu\text{m}$\cite{PizzNATC2016}, known as `blisters'\cite{PizzNATC2016} or `bubbles'\cite{MayoNL2011,WangSCI2013,HaigNATM2012}, which form due to the interplay of LM elastic properties and van der Waals forces\cite{KhesNATC2016}.

This aggregation of contaminants into blisters leaves the regions located between them with clean interfaces\cite{HaigNATM2012}, and devices can therefore be fabricated exploiting these areas\cite{KretNL2014}. However, the device size is constrained by the blister spacing, typically$\sim$1-10$\mu$m\cite{KretNL2014}. For hBN encapsulated SLG, this also limits fundamental studies, since at cryogenic temperatures (T) the charge carrier mean free path in SLG, $l_{m}$, can reach$>10\mu\text{m}$\cite{BansNL2016,WangSCI2013}, which results in the electrical conductivity becoming limited by scattering of carriers with device edges, due to the lateral dimensions of the sample, W, being$<l_{m}$\cite{MayoNL2011}. In general, the random placement of blisters is incompatible with scalable fabrication processes, requiring devices to be deterministically located. Thus, it is a must to develop methods to eliminate blisters.

Blister-free areas$>10\mu\text{m}$ can be obtained by using a `hot pick-up' technique\cite{PizzNATC2016}, where interfacial contaminations are removed during encapsulation by bringing the layers together in a conformal manner at$\sim$110$^\circ$C\cite{PizzNATC2016}. The cleaning in this process is attributed to the higher diffusivity of the contaminants at $110^\circ$C than at RT\cite{PizzNATC2016}, allowing them to diffuse out of the sample during encapsulation. Large blister-free regions were also reported in Ref.\citenum{WangSCI2013}, up to$\sim300\mu\text{m}^{2}$, although no explanation of how blisters are avoided was given. In Refs.\citenum{PizzNATC2016},\citenum{WangSCI2013} residual blisters sometimes remained within the samples due to incomplete cleaning during transfer. Furthermore, the technique of Ref.\cite{PizzNATC2016} only achieves clean interfaces when the encapsulation is performed slowly, with lateral speeds$<1\mu\text{m}/$s, limiting its applicability to scalable processes. The required time would increase as the total number of interfaces within the LMH increases. Ref.\citenum{PizzNATC2016} also produced LMHs using SLG intentionally contaminated with PMMA residuals left from lithographic processing, suggesting that the `hot pick-up' technique could be used to exclude these polymer residuals. However, no comparison was given of $\mu$ of samples produced using clean and polymer contaminated SLG.

Here we show how to remove contaminations trapped within already assembled LMHs. This is achieved by laminating the LMH onto a SiO$_{2}$ substrate at$\sim$180$^\circ$C. At this T the blisters become physically mobile, enabling them (and the contaminants trapped inside) to be pushed to the sample edges, where they are eliminated. We achieve blister-free hBN-encapsulated SLG with lateral dimensions up to$\sim$5000$\mu$m$^2$, limited only by the size of the exfoliated flakes, one order of magnitude larger than Refs.\citenum{PizzNATC2016,WangSCI2013}. We manipulate blisters at speeds$>10\mu\text{m}\slash\text{s}$, over an order of magnitude faster than Ref.\citenum{PizzNATC2016}. Our approach also allows the LMH interfaces to be cleaned simultaneously, unlike the `hot pick-up', where the interfaces need to be cleaned sequentially\cite{PizzNATC2016}. Thus, our total cleaning time is independent on the number of layers. We also demonstrate blister cleaning of LMHs with MoS$_{2}$, indicating the general suitability of our approach. We fabricate 4 terminal devices, exploiting the entire LMHs area, with Hall bar widths up to $W=24\mu$m. We achieve $\mu$ up to 180000cm$^2\ V^{-1}s^{-1}$ at RT. $\mu$ is consistently high across all samples, with an average$\sim160000$cm$^2$ V$^{-1}$s$^{-1}$ across 15 Hall bars. Hall bars with $\mu$ up to$\sim 150000$cm$^2$ V$^{-1}$s$^{-1}$ were previously reported\cite{MayoNL2011,KretNL2014}, but in devices with $W<3\mu$m, limited by the blister spacing, while our technique enables $\mu>100000$cm$^{2}$ V$^{-1}$s$^{-1}$ to be consistently achieved over the entire LMH dimensions. We report $\mu$ up to$\sim$1.8$\times$10$^6$cm$^2\ V^{-1}s^{-1}$ at 9K. To the best of our knowledge, this is the highest diffusively measured $\mu$ in SLG, with similar values only obtained using ballistic measurements where $l_{m}$ exceeded the sample dimensions\cite{WangSCI2013,BansNL2016}. We also show that our approach works on SLG intentionally exposed to PMMA, acetone and isopropyl alcohol (IPA) before encapsulation, achieving $\mu$ up to$\sim$150000cm$^2\ V^{-1}s^{-1}$ at RT after cleaning, i.e., there is no $\mu$ degradation compared to non-contaminated SLG. This is significant because, to date, exposure of SLG to polymers and solvents before encapsulation resulted in samples with limited $\mu$, attributed to the contaminants trapped at the SLG-hBN interface\cite{WangSCI2013,BansSCIA2015,MayoNL2011,KretNL2014}. E.g. Ref.\citenum{PetrACSN2015} used a semi-dry transfer and achieved $\mu\sim50000$cm$^2$V$^{-1}$s$^{-1}$ at 1.6K, a factor$\sim35$ lower than us. Ref.\citenum{DeanNATN2010} used non-encapsulated SLG on a bottom hBN, with SLG exposed to both polymers and solvents, followed by annealing at 300$^{\circ}$C to remove residuals, achieving $\mu \lesssim 30000$cm$^2$ V$^{-1}$s$^{-1}$ at charge carrier concentrations $n>1\times10^{12}$cm$^{-2}$ and 7.2K, a factor$\sim40$ lower than us. In Ref.\citenum{KretNL2014} SLG exposed to polymer and solvents before encapsulation showed $\mu$ up to$\sim 150000$cm$^2$ V$^{-1}$s$^{-1}$ at T$<$10K, one order of magnitude lower than us, with device size still limited by residual blisters\cite{KretNL2014}. Thus, our approach paves the way to the optimization of scalable techniques, such as wet\cite{SukACSN2011,BaeNATN2010} and (or) polymer assisted transfer\cite{CaldEST2010,MattJMC2011,GaoNATC2012,WangACSN2011}, for the fabrication process of high $\mu$ encapsulated LMHs.
\section{Results and discussion}
\subsection{Encapsulation, cleaning, and device fabrication}
\begin{figure*}
\centerline{\includegraphics[width=180mm]{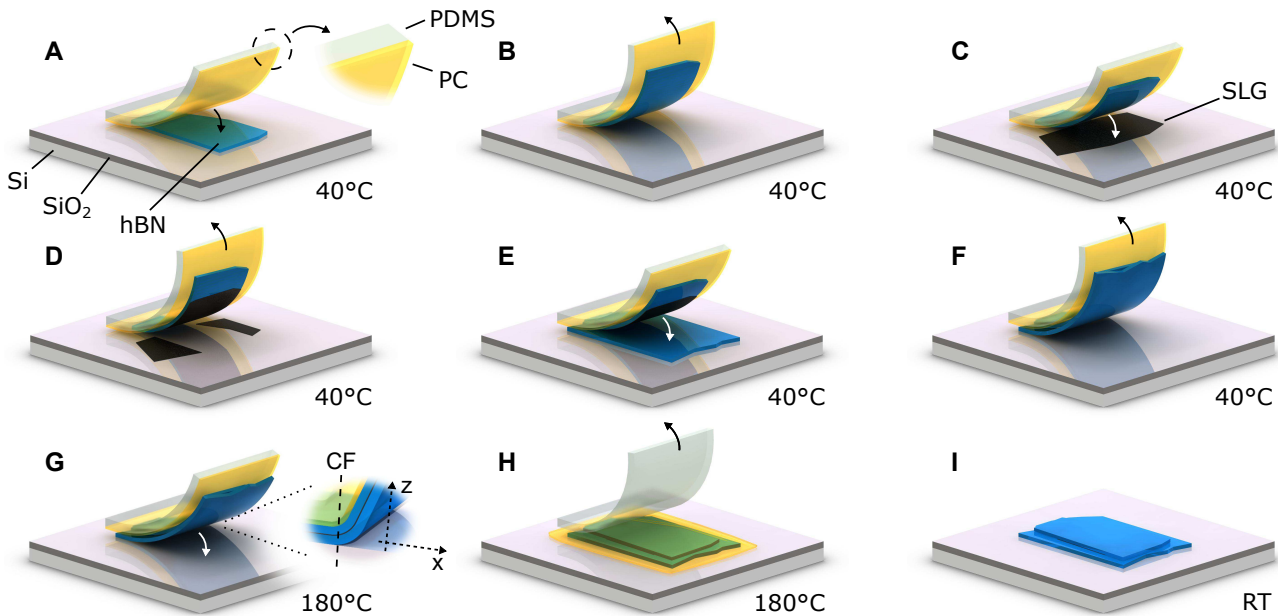}}
\caption{Scheme of the cleaning process. a) A stamp, consisting of a PC film (yellow) mounted on a PDMS block (white-translucent) is brought into contact with a hBN flake (blue) exfoliated on SiO$_{2}$+Si (purple/light grey). b) The stamp is withdrawn, picking up the hBN. c) The hBN is lowered into contact with an exfoliated SLG (black), and then withdrawn, d), picking the SLG portion in contact with hBN. e) hBN and SLG are brought into contact with another hBN flake, forming the encapsulated stack. f) The encapsulated stack is picked up from the SiO$_{2}$+Si substrate. Steps a)-f) are performed at $40^{\circ}$C. g) T is raised to $180^{\circ}$C and the encapsulated stack is laminated onto SiO$_{2}$+Si. The CF is defined as the interface between the LMH suspended portion and that in contact with the SiO$_{2}$+Si. Control over the stamp height (z) determines the CF lateral movement (x). This is achieved by tilting the PDMS block, such that the stamp first contacts the substrate on one side. As the CF encounters blisters, these are manipulated and removed. h) The stamp is withdrawn. The PC adheres to the substrate, PDMS is peeled away. i) PC is dissolved in chloroform.}
\label{fig:Fig1}
\end{figure*}
Fig.\ref{fig:Fig1} shows a schematic representation of our approach to produce hBN/SLG/hBN LMHs. Flakes of hBN and SLG are prepared by micro-mechanical cleavage (MC)\cite{NovoPNAS2005} on Si+285nm SiO$_2$ from bulk crystals of graphite (NGS Naturgraphit) and hBN. hBN single crystals are grown under high pressure and high temperature, as detailed in Ref.\citenum{TaniJCG2007}. The graphite is first cleaved using adhesive tape. The Si+SiO$_2$ substrate is then exposed to an oxygen plasma (100W, 360s). The surface of the tape is brought into contact with the SiO$_{2}$ substrate, which is then placed on a hot plate at 100$^\circ$C for 2mins, before the tape is removed. Heating the substrate allows us to achieve large ($>100\mu$m) SLG flakes, whereas flakes produces without heating are typically $<50\mu$m in size, in agreement with findings of Ref.\citenum{HuanASCN2015}. For the exfoliation of hBN, no plasma treatment of the SiO$_{2}$ surface is used, as we find this has no effect on the flakes lateral size.

SLG and hBN flakes are identified prior to transfer by a combination of optical microscopy\cite{CasiNL2007} and Raman spectroscopy\cite{FerrPRL2006,FerrNN2013,ReicPRB2005,NemaPRB1979,ArenNL2006}. To showcase the effectiveness of our approach, we fabricate LMHs with a range of hBN thicknesses, $t_{hBN}$ (2-300nm), and widths, $W_{hBN}$ (up to$\sim200\mu$m), observing blister manipulation and cleaning in all cases. For devices targeting $\mu>100000$cm$^{2}$V$^{-1}$s$^{-1}$, we select bottom hBN flakes with $t_{hBN}>10$nm, as below this value the hBN no longer screens the roughness of the underlying SiO$_{2}$\cite{DeanNATN2010}. Furthermore, thin ($t_{hBN}<10$nm) hBN flakes may provide insufficient screening from charged impurities in the underlying SiO$_{2}$\cite{HongPRB2009,BursNL2013}, and both of these factors may degrade the SLG $\mu$. We note that, to the best of our knowledge, all $\mu \gtrsim 100000$cm$^{2}$V$^{-1}$s$^{-1}$ samples in literature used $t_{hBN}\gtrsim10$nm\cite{DeanNATN2010,WangSCI2013,KretNL2014,BansSCIA2015,MayoNL2011}.
\begin{figure*}
\centerline{\includegraphics[width=180mm]{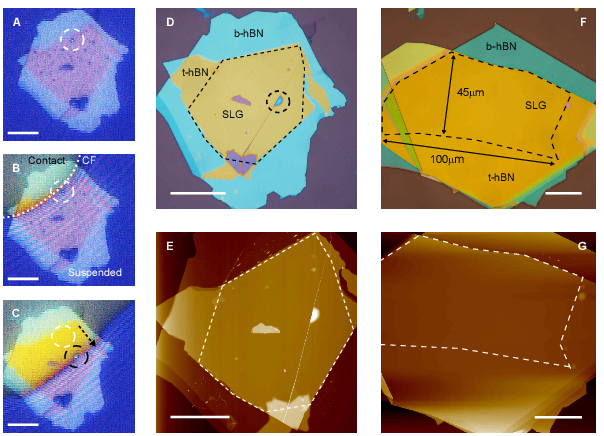}}
\caption{Optical images and AFM scans of LMH cleaning. a) Encapsulated sample suspended on the PC stamp above Si+SiO$_{2}$. One blister is highlighted with a dashed white circle. Other blisters are seen as dark spots. b) Optical image as the sample is laminated onto Si+SiO$_{2}$. The CF between PC and substrate is marked with a white dashed line. Above the CF, the PC is in contact with SiO$_{2}$, while below it is suspended. c) As the CF advances it pushes the blisters. The blister in b), originally in the position marked by the white circle, has now moved, as marked by the black circle. The arrow shows the direction of movement. d) Optical image of sample a) after cleaning. One blister remains, pinned by a hBN wrinkle, highlighted by the dashed circle. t-hBN: top hBN. b-hBN: bottom hBN. e) AFM scan of the sample in d). f,g) Optical image and AFM scan of a sample with a blister-free area of encapsulated SLG. The dashed lines mark the SLG. Scale bars 20$\mu$m.}
\label{fig:Fig2}
\end{figure*}
\begin{figure*}
\centerline{\includegraphics[width=180mm]{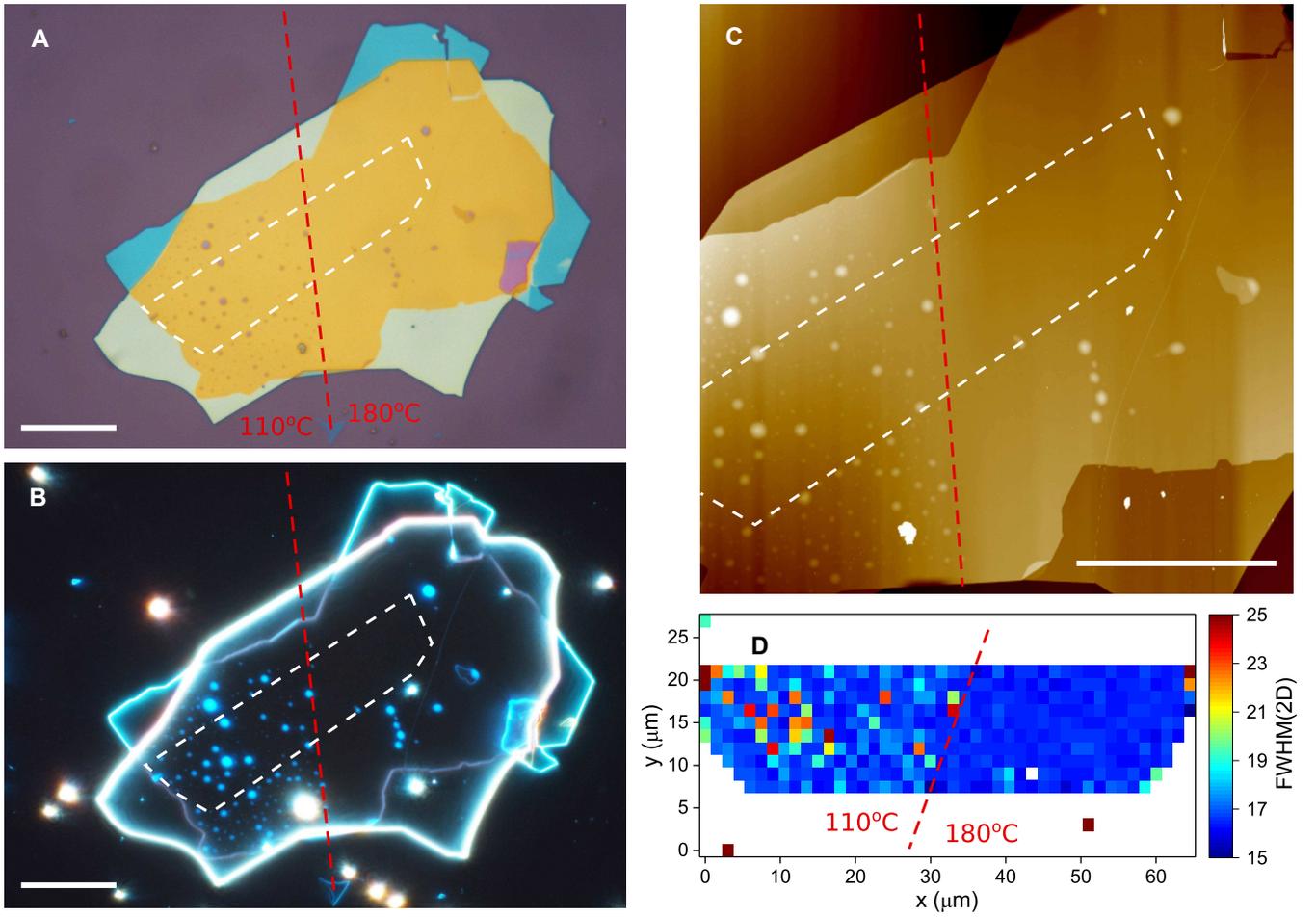}}
\caption{Effect of the T on blister cleaning. a) Optical image of a sample partially cleaned at 110 and 180$^{\circ}$C. The interface between the two regions is marked by a dashed red line. The SLG location is marked by a while dashed line. b,c) optical dark field and AFM images of the same sample. d) FWHM(2D) map across the sample. Scale bars 20$\mu$m.}
\label{fig:Fig4}
\end{figure*}
\begin{figure*}
\centerline{\includegraphics[width=180mm]{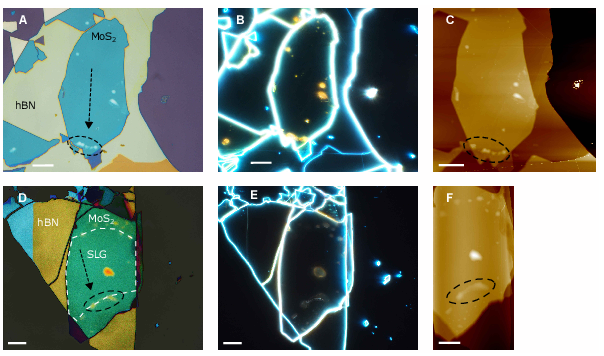}}
\caption{Blister manipulation in LMHs containing MoS$_{2}$. a-c) Bright field, dark field, and AFM images of MoS$_{2}$/hBN after cleaning. d-f) Bright field, dark field, and AFM images of hBN/SLG/MoS$_{2}$ after cleaning. In a,c,d,f the black dashed circles mark where the majority of blisters have been pushed. The arrows indicate the direction of the blister cleaning. Scale bars 10$\mu$m.}
\label{fig:Fig3}
\end{figure*}

In order to pick up and transfer the flakes we use a stamp consisting of a layer of polycarbonate (PC) mounted on a block of polydimethylsiloxane (PDMS) for mechanical support, Fig.\ref{fig:Fig1}a. The stamp is similar to that suggested in Ref.\citenum{WangSCI2013}, however PC is used instead of poly-propylene carbonate (PCC) as our cleaning requires T$\sim$180$^{\circ}$C, well above the PCC glass transition $T_g\sim40^{\circ}$C\cite{LuinAPS245}. The PC is prepared by drop casting a solution in chloroform ($5\%$ weight) onto a glass slide. After drying, the resultant film is picked up and mounted on a PDMS block ($\sim$3mm thick), which is then placed on a glass slide attached to a micro-manipulator (resolution$\sim1\mu\text{m}$) under a microscope. The Si+SiO$_2$ substrates, with the flakes to be transferred, are positioned underneath the micro-manipulator, on a heated stage, enabling T control from RT up to$\sim300^{\circ}\text{C}$.

The process begins by placing the PC into contact with a selected hBN flake, then withdrawing, while keeping the substrate at 40$^\circ$C. This T is chosen because it allows us to pick both hBN and SLG flakes with a success rate of 100$\%$ (as compared to RT, where this is $<90\%$). The hBN adheres to the PC surface and is removed from the Si+SiO$_{2}$ as the stamp is lifted, Fig.\ref{fig:Fig1}b. We then position the hBN over a chosen SLG flake and bring the two into contact, before again withdrawing while still at 40$^\circ$C. The portion of the SLG in contact with hBN delaminates from the Si+SiO$_{2}$, while that in contact with the PC remains on the Si+SiO$_{2}$, Fig.\ref{fig:Fig1}c,d. This is attributed to the preferential adhesion of SLG to hBN\cite{PizzNATC2016}. hBN and SLG flakes are then aligned and brought into contact with another (bottom) hBN flake, Fig.\ref{fig:Fig1}e, encapsulating the SLG.

We next withdraw the stamp with the LMH still attached to the PC, suspending it above the Si+SiO$_{2}$, Fig.\ref{fig:Fig1}f. The stage T is increased to 180$^\circ$C, following which the stamp is brought into contact with the substrate, Fig.\ref{fig:Fig1}g. During this step the PDMS block is tilted$\sim1^{\circ}$, so that contact with the substrate first occurs on one side of the stamp, and then advances horizontally across it. Control over the stamp vertical position also defines the position of the contract front (CF) in the horizontal direction. The CF is the interface between the portion of the stamp in contact with Si+SiO$_{2}$, and that suspended, as in Fig.\ref{fig:Fig1}g. At T=180$^\circ$C the PC is above $T_{g}\sim$150$^\circ$C\cite{FanMTS1997}, resulting in decreased viscosity\cite{YangPES1997}, allowing greater control over its lateral movement. Below $T_{g}$, the CF can move laterally in uncontrolled, discrete jumps.

As the CF approaches the encapsulated SLG, we observe the aggregation of numerous blisters, Fig.\ref{fig:Fig2}a. The typical blister coverage can also be seen in the un-cleaned portion of the sample in Fig.\ref{fig:Fig4}. We attribute this to the LMH approaching the Si+SiO$_{2}$ surface, resulting in its T increasing to$\sim180^{\circ}\text{C}$. At RT, trapped contaminants cover the sample interfaces\cite{PizzNATC2016}, but become increasingly mobile, segregating into spatially localized blisters as the T rises above$\sim70^{\circ}\text{C}$\cite{PizzNATC2016}.

When the CF passes across the encapsulated stack, the stack is laminated onto the Si+SiO$_{2}$, Fig.\ref{fig:Fig1}g. This pushes any blisters within the LMH in the direction of the advancing CF. As blisters are swept through the LMH they collide and aggregate. They continue to move until they reach the LMH edge, at which point they are eliminated, or until they reach a physical discontinuity, such as a crack or wrinkle in the hBN or SLG, which may pin them. Once the CF has fully passed across the encapsulated stack and the blister removal is complete, the stamp is withdrawn, Fig.\ref{fig:Fig1}h. At 180$^\circ$C the PC preferentially adheres to the SiO$_{2}$, allowing the PDMS to be peeled away, leaving the PC adhered to the SiO$_{2}$+Si surface, Fig.\ref{fig:Fig1}h. The PC is then removed by rinsing the sample in chloroform for $\sim10$ minutes, Fig.\ref{fig:Fig1}i.

Figs.\ref{fig:Fig2}a-c show the movement of blisters in response to the advancing CF. Fig.\ref{fig:Fig2}a is the sample before cleaning, suspended on the PC stamp above Si+SiO$_{2}$. Numerous blisters can be seen. In Fig.\ref{fig:Fig2}b the CF (marked by the white dashed line) is advancing across the LMH. Above the CF (yellow optical contrast) the PC is in contact with Si+SiO$_{2}$. In Fig.\ref{fig:Fig2}c the CF has advanced further. One blister is highlighted, with its initial location marked by a dashed white circle in Figs.\ref{fig:Fig2}a-c, and by a dashed black circle in Fig.\ref{fig:Fig2}c after being moved by the advancing CF. Fig.\ref{fig:Fig2}d is the same sample after cleaning, with an atomic force microscope (AFM, Bruker Dimension Icon, operated in peak force) scan in Fig.\ref{fig:Fig2}e. One blister (highlighted by a dashed black circle) remains, pinned by a wrinkle. A second LMH also encapsulated and cleaned using the same method is shown in Fig.\ref{fig:Fig2}f, with AFM in Fig.\ref{fig:Fig2}g. This is blister-free over$\sim$100$\times$45$\mu$m.

Blisters can be manipulated at speeds$>10\mu$m/s. They can also be pulled by withdrawing instead of advancing the CF, i.e., they can be continuously manipulated both forwards and backwards. The presence of SLG in the LMH plays a significant role in the ability to manipulate the blisters using the CF. While blisters are always pushed by the CF in the hBN/SLG/hBN portion of the LMH, they can become immobile at the hBN/hBN interface. They can also be pinned at the SLG edge (see the right-hand edge of the dashed white line in Fig.\ref{fig:Fig2}g). This can result in samples where the SLG region is blister-free, but surrounded at the edges by blisters.

Our cleaning method also works in LMHs based on different materials, such as hBN/MoS$_{2}$ and hBN/SLG/MoS$_{2}$. Figs.\ref{fig:Fig3}a-c show a hBN/MoS$_{2}$ sample in which blister cleaning is performed. The direction of cleaning is indicated by the dashed arrow. The majority of blisters have been pushed to the LMH edge, as highlighted by a dashed circles in a and c. Figs.\ref{fig:Fig3}d-f are hBN/SLG/MoS$_{2}$ LMHs. The majority of the blisters have been pushed to the edge of the SLG, as indicated by the dashed circles in d and f.

Ref.\onlinecite{PizzNATC2016} reported that T plays a key role in the ability to exclude contaminants from LMH interfaces. Thus, we now consider the effectiveness of blister manipulation at $110^{\circ}$C and $180^{\circ}$C. In the cleaning step, we initiate the process at $110^{\circ}$C, until the CF has passed half way across the sample. Following this, T is raised to $180^{\circ}$C and the CF is advanced over the remaining portion of the LMH. Figs.\ref{fig:Fig4}a-c are optical bright field and dark field images and an AFM scan of the sample. In the portion of the LMH cleaned at $110^{\circ}$C numerous blisters can be observed, at $180^{\circ}$C this appears blister-free. This demonstrates the effect of T on the cleaning process. At $110^{\circ}$C the physical mobility of the blisters is insufficient for them to be manipulated, while at $180^{\circ}$C they are mobile and can be removed from the LMH.

To further understand the effect of T we consider a model based on quantized fracture mechanics\cite{PugnPM84}. In a stack formed by PDMS, PC, hBN, SLG and hBN, laminated onto SiO$_2$ (as in Fig.\ref{fig:Fig1}g), we can evaluate the elastic energy per unit length stored in the LMH around the zone of separation from the substrate (i.e. the curved region in Fig.\ref{fig:Fig1}g). This can be written as: $\frac{\mathrm d L}{\mathrm d s}=\frac{1}{2}\frac{EI}{R^{2}}$, where $R$ is the radius of curvature of this zone and $EI$ is the LHM rigidity (i.e. the Young's modulus multiplied by the moment of inertia of the cross-section of the layer, in N$\times$m$^2$). Considering the 5 materials in the stack (PDMS, PC, hBN, SLG and hBN), each with Young's moduli $E_i$ and thickness $h_i$, we first derive the position of the elastic neutral axis (i.e. where the stresses are 0) as for Ref.\citenum{CarpFC5}: $y_0=\frac{\sum_{i=1}^{N} E_ih_iy_i}{\sum_{i=1}^{N} E_ih_i}$, where $y_i$ are the positions of the barycenters of each layers. We get $EI=\frac{b}{12}\sum_{i=1}^{N}[E_ih_i^3+12E_ih_i(y_i-y_0)^2]$\cite{CarpFC5}, where $b$ is the width of the stack. For a homogeneous layer with Young's modulus $E$ and total thickness $h=\sum_{i=1}^{N}h_i$, we have $EI=E\frac{bh^3}{12}$\cite{KeSM438}, where $I$ [m$^4$] is the momentum of inertia of the layer. Equating the last two expressions we get the homogenized Young's modulus $E_{homog}$ of the stack. During adhesion, the energy balance imposes $\frac{\mathrm d L}{\mathrm d s}=2\Gamma b$\cite{PugnPM84}, where $\Gamma$ is the adhesion energy (in J/m$^2$) between the stack and the substrate. The pressure generated at the interface is thus $p\cong \frac{\Gamma}{R}=4\sqrt{\frac{3\Gamma^{3}}{Eh^{3}}}$\cite{PugnPM84}. The pressure inside a circular-shaped blister of radius $a$ needed for its propagation is $p_{c}\cong \sqrt{\frac{2\alpha\gamma_{i,j} E}{\pi (a+q/2)}}$\cite{HuanASCN2015}, where $\gamma_{i,j}$ is the adhesion energy between two layers i,j (i.e. hBN and SLG) forming the blister, $\alpha$ is a non-dimensional shape factor close to unity\cite{PugnPM84}, and $q$ is the minimum value of blister advancement. The condition for blister cleaning is $p>p_{c}$. Noting that the adhesion energies are T-dependent and present maximal values at a given T (e.g., SLG's adhesion to SiO$_2$ is maximum at$\sim250^{\circ}C$\cite{HeJPCC119}), we get $\Gamma=\Gamma^{(max)}f\left(\mathrm{T}\right)$ and $\gamma_{i,j}=\gamma^{(max)}_{i,j} g\left(\mathrm{T}\right)$, where $0<f\left(\mathrm{T}\right)$, $g(\mathrm{T})\leq 1$. Similarly, $E_i (\mathrm{T})$ are T-dependent, thus $E(\mathrm{T})=E^{(max)}e(\mathrm{T})$, where $0<e(\mathrm{T})\leq1$. Accordingly, for blister cleaning the following condition must be satisfied:
\begin{equation}
C(\mathrm{T}) = \frac{f(\mathrm{T})^3}{e(\mathrm{T})^2g(\mathrm{T})}>\frac{\alpha E^{(max)2}h^3\gamma^{(max)}_{i,j}}{24 \pi (a+q/2)\Gamma^{(max)3}}=A
\end{equation}
\begin{figure*}
\centerline{\includegraphics[width=180mm]{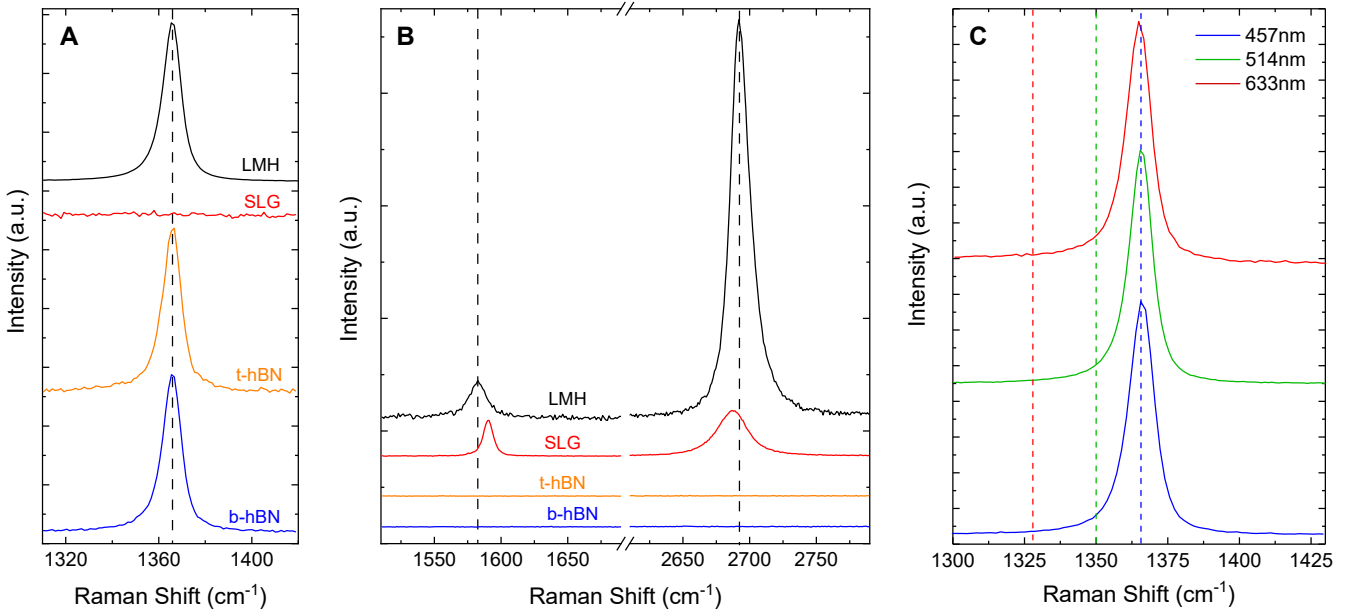}}
\caption{Raman spectra of LMs before and after assembly, taken at 514nm. The bottom hBN flake (b-hBN) is shown in blue, the top hBN flake (t-hBN) orange, the SLG flake in red, and the assembled LMH in black. a) hBN E$_{2g}$ spectral region. The measured spectra are all normalised to the height of the E$_{2g}$ peak for clarity. b) G and 2D peak spectral region. The spectra containing SLG peaks are normalised to the height of the G peak for clarity. The spectra associated with t-hBN and b-hBN have the same scaling as in a. The E$_{2g}$, G and 2D peaks after encapsulation are marked by dashed black lines. c) Raman spectra measured at 457, 514, and 633nm. The expected Pos(D) at 457, 514, and 633nm are shown by dashed lines in blue, green and red respectively. The spectra are normalised to the height of the 2D peak.}
\label{fig:Fig5}
\end{figure*}
where we introduced the dimensionless cleaning thermal driving force $C$(T) and the blister resistance $A$. By increasing T we can simultaneously increase $C$(T) and decrease $A$, e.g. by reducing $E^{(max)}$ imposing a glass transition of a polymer layer. Thus, in our case, well above the PC T$_{g}$, $E_{PC}$ becomes negligible. For perfect cleaning $a=0$ and $A$ is maximal. Considering $f(\mathrm{T})\cong g(\mathrm{T})$  (same T dependence of $\gamma_{i,j}$ and $\Gamma$) and $e(\mathrm{T})\cong 1$ (nearly T-independent homogenized $E$), the blister cleaning requires T in the range $\mathrm{T}_0-\Delta \mathrm{T}_- \leq \mathrm{T} \leq \mathrm{T}_0 -\Delta \mathrm{T}_+$, where $\mathrm{T}_0$ is the T at which surface energies are maximal, i.e. $f(\mathrm{T}_0 )=g(\mathrm{T}_0 )=1$  (note that $\Delta \mathrm{T}_-=\Delta \mathrm{T}_+$ if a symmetric function is assumed). In this case, the condition for blister cleaning becomes $C(\mathrm{T})\cong g(\mathrm{T})^2>A$. Considering the T dependence of the adhesion energy for SLG on SiO$_2$\cite{HeJPCC119}, we can assume $\mathrm{T}_0 \cong 250 ^\circ$ C. Noting that for PC, $\mathrm{T}_g \cong 150 ^\circ $C, we expect a 150-250$^\circ$C range of minimal T for blister cleaning, in good agreement with our observation of no blister cleaning below 150$^\circ$C and good cleaning at 180$^\circ$C.
\begin{figure*}
\centerline{\includegraphics[width=180mm]{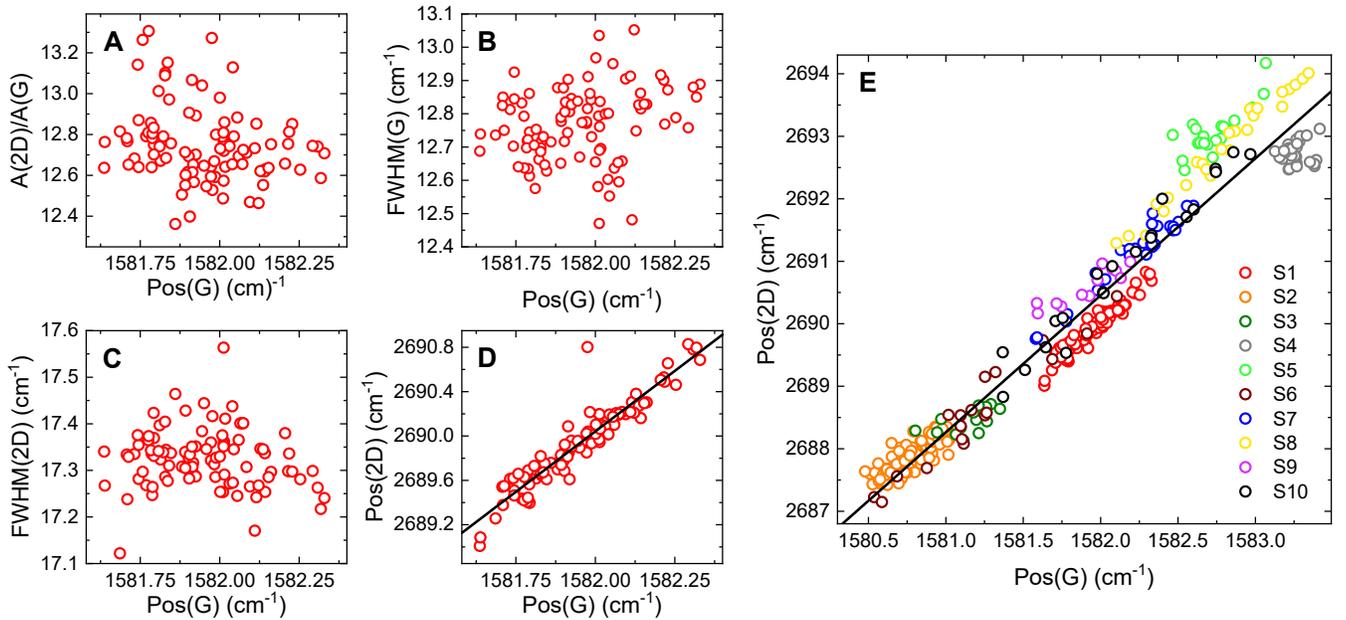}}
\caption{a)-d) Pos(2D), A(2D)\slash A(G), FWHM(2D), FWHM(G) all vs. Pos(G) mapped across a $20\mu$m$\times20\mu$m region of a Hall bar (optical image shown in the inset of Fig. \ref{fig:Fig7}a). e) Pos(2D) vs. Pos(G) for ten (S1-S10) different samples. S1 corresponds to the measurements in a-d. Solid black lines represent linear fits to the data.}
\label{fig:Fig6}
\end{figure*}

The quality of the flakes is monitored both before and after assembly by Raman spectroscopy. Measurements are performed using a Renishaw inVia at 457, 514 and 633nm. Fig.\ref{fig:Fig5}a-c plots the spectra of a typical sample, with 92 and 176nm top and bottom hBN flakes. Fig.\ref{fig:Fig5}a shows that the $E_{2g}$ peak for both the bottom and top hBN are at 1366cm$^{-1}$, with full-width half maximum (FWHM)$\sim$9.2cm$^{-1}$ and 8.6cm$^{-1}$, respectively, as expected for bulk hBN\cite{ArenNL2006,ReicPRB2005,NemaPRB1981}. The SLG G and 2D peaks before transfer are plotted in Fig.\ref{fig:Fig5}b. The 2D peak can be fit with a single Lorentzian, with a full width half maximum FWHM(2D)$\sim$26cm$^{-1}$, and position Pos(2D)$\sim$2687cm$^{-1}$, as expected for SLG\cite{FerrPRL2006,FerrNN2013}. No D peak can be seen, indicating negligible defects\cite{FerrPRL2006,FerrNN2013,CancNL2011}. The position of the G peak Pos(G)$\sim$1590cm$^{-1}$, the full-width half maximum of the G peak FWHM(G)$\sim$8cm$^{-1}$, and the intensity and areas ratio of the 2D and G peaks I(2D)$\slash$I(G)$\sim$1.3, A(2D)$\slash$A(G)$\sim3.9$ indicate the sample is doped with E$_{F}\gtrsim300$meV\cite{DasNN2008,BaskPRB2009}. The LMH spectrum is shown in black in Fig.\ref{fig:Fig5}b. The hBN E$_{2g}$ peak is now a combination of those of both top and bottom hBN. This yields a single peak with Pos(E$_{2g}$)$\sim$1366cm$^{-1}$ and FWHM(E$_{2g}$)$\sim9.3$, as expected considering both flakes are bulk\cite{ArenNL2006,ReicPRB2005,NemaPRB1981}. For the encapsulated SLG we have Pos(2D)$\sim$2693cm$^{-1}$, Pos(G)$\sim$1583cm$^{-1}$, FWHM(G)$\sim15$cm$^{-1}$, I(2D)/I(G)$\sim11.4$cm$^{-1}$ and A(2D)/A(G)$\sim 12.9$cm$^{-1}$, indicating E$_{F}\ll100$meV\cite{DasNN2008,BaskPRB2009}. FWHM(2D) decreases to$\sim17$cm$^{-1}$ after encapsulation, indicating a reduction in the nanometer-scale strain variations within the sample\cite{NeumNATC2015,PisaNM6,YanPRL98}. We note that the E$_{2g}$ peak of hBN may overlap the D peak. This is a general issue in hBN-encapsulated samples. However, the D peak shifts with excitation energy by$\sim50$cm$^{-1}\slash$eV\cite{VidaSSC39,PocsJNCS227} due to a combination of its double resonance activation\cite{ReicPRB2005,BaraOS62} and a Kohn Anomaly at the K point of the Brillouin Zone\cite{PiscPRL93}, while the E$_{2g}$ of hBN does not, since hBN has a band gap and no Kohn anomalies nor double resonances are present\cite{ReicPRB2005,PiscPRL93}. Fig.\ref{fig:Fig5}c compares the Raman spectra at 457, 514 and 633nm. No D peak is seen even at 633nm, where it should be well clear of the E$_{2g}$ of hBN, thus ensuring no extra defects are introduced in SLG by the transfer and cleaning processes.

Following encapsulation and blister removal, we process our LMHs into Hall-bars for 4-terminal transport measurements. The LMH is first dry etched, defining the geometry, as well as exposing the SLG edge. Depositing metal onto the exposed edges results in ohmic contacts between the SLG and metal\cite{WangSCI2013}. We first deposit an Al mask using e-beam lithography, metal evaporation and lift-off. Dry etching is performed with a forward power of 20W in a plasma formed from a mixture of CF$_4$ and O$_2$. After wet-etching to remove the Al mask, metal contacts are patterned by e-beam lithography followed by either e-beam evaporation and lift-off of 5/150nm Cr/Au\cite{WangSCI2013}, or DC sputtering and lift-off of 5/150nm of Cr/Cu. We fabricate Hall bars with W up to 24$\mu$m, one order of magnitude larger than the W$\sim$1-3$\mu$m commonly reported in literature\cite{KretNL2014,BansSCIA2015,MayoNL2011}.

\begin{figure}
\centerline{\includegraphics[width=80mm]{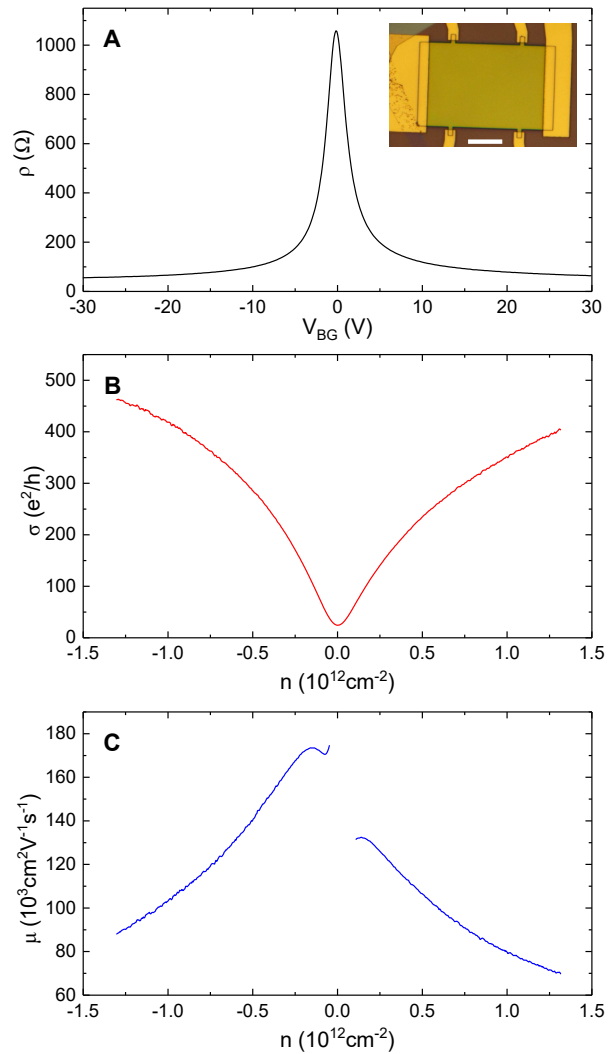}}
\caption{4 terminal measurements at RT for sample S1. a) Resistivity vs. gate voltage. Inset: optical image of the encapsulated SLG Hall bar. W=24$\mu\text{m}$. Scale bar 10$\mu\text{m}$. b) Conductivity vs. charge carrier density. c) Mobility as a function of charge carrier density.}
\label{fig:Fig7}
\end{figure}

We then perform Raman mapping after device fabrication. The data in Fig.\ref{fig:Fig6}a-d are taken from a$\sim 20\mu\text{m}\times 20\mu$m map on the Hall bar in the inset in Fig.\ref{fig:Fig7}a. Pos(G) is sensitive to both doping\cite{DasNN2008} and stain\cite{MohiCMMP2009}, meaning that local variations in these quantities manifest as a spread in the G peak position, i.e., $\Delta$Pos(G). From Figs.\ref{fig:Fig6}a-d $\Delta$Pos(G)$\sim 0.6$cm$^{-1}$. Figs.\ref{fig:Fig6}a,b plot A(2D)/A(G) and FWHM(G) as a function of Pos(G), showing no correlation. This indicates that the contribution to $\Delta$Pos(G) due to doping is negligible\cite{CasiAPL2007,PisaNM6,DasNN2008} and that the trend in Fig.6d is due to strain ($\epsilon$). Fig.\ref{fig:Fig6}d plots Pos(2D) as a function of Pos(G). A linear correlation can be seen with slope $\Delta$Pos(2D)/$\Delta$Pos(G)$\sim2.18$. A similar trend was reported in Ref.\citenum{LeeNATC2012}, with a slope$\sim2.2$.

The rate of change of Pos(2D) and Pos(G) with strain is ruled by the Gr\"{u}neisen parameters ($\gamma$)\cite{MohiCMMP2009}, which relate the relative change in the peak positions in response to strain, i.e. $\gamma =[\omega - \omega_{0}]/[2\epsilon\omega_{0}]$, where $\omega$ is the frequency of Raman peaks at finite strain and $\omega_{0}$ the frequency at zero strain\cite{MohiCMMP2009}. For biaxial strain the Gr\"{u}neisen parameters for G and 2D peak are respectively $\gamma_{G}\sim1.8$ and $\gamma_{2D}\sim2.6$, resulting in $\Delta$Pos(2D)/$\Delta$Pos(G)$\sim2.5$\cite{MohiCMMP2009,ZabeNL2012,ProcCMMP2008}. In the case of uniaxial strain $\gamma_{G}\sim1.8$\cite{MohiCMMP2009}, however extraction of $\gamma_{2D}$ is not straightforward, as uniaxial strain shifts the relative position of the Dirac cones in SLGs band structure\cite{MohiCMMP2009,ZabeNL2012}, which in turn effects the 2D peak as it is activated by an intervalley scattering process\cite{FerrNN2013,MohiCMMP2009}. Ref.\cite{MohiCMMP2009} extracted from measurements an upper bound $\gamma_{2D}\sim3.55$ and theoretically calculated $\gamma_{2D}\sim2.7$, consistent with experimentally reported $\Delta$Pos(2D)/$\Delta$Pos(G)$\sim2-3$\cite{MohiCMMP2009,YoonPRL2011,LeeNATC2012}. Biaxial strain can be differentiated from uniaxial from the absence of G and 2D peak splitting with increasing strain\cite{FerrNN2013}, however at low ($\lesssim0.5\%$) strain the splitting cannot be resolved\cite{MohiCMMP2009,YoonPRL2011}. Due to these factors the presence (or coexistence) of biaxial strain cannot be ruled out in our samples. For uniaxial(biaxial) strain, Pos(G) shifts by $\Delta $Pos(G)$\slash\Delta \epsilon\sim 23(60) \text{cm}^{-1}\slash \%$ \cite{MohiCMMP2009,YoonPRL2011,ZabeNL2012}. For intrinsic SLG (E$_{F}\ll 100$meV), the unstrained value of Pos(G)$\sim 1581.5$cm$^{-1}$ for 514nm excitation\cite{FerrPRL2006}. For the sample in Fig.\ref{fig:Fig6}d, $\Delta$Pos(G)$\sim0.6$cm$^{-1}$ equates to $\Delta \epsilon \lesssim0.026\%$. The average Pos(G)$\sim 1582$cm$^{-1}$ indicates an average strain $\epsilon\sim 0.025\%$. Fig.\ref{fig:Fig6}e plots Pos(2D) as a function of Pos(G) for 10 samples (S1-S10), all prepared using the blister cleaning method detailed earlier using $t_{hBN}>10$nm. The data shows a linear trend, with a slope$\sim2.19$. $\Delta$Pos(G) ranges from$\sim$0.5 to 2cm$^{-1}$, indicating differences in $\Delta\epsilon$ up to a factor$\sim4$. The average Pos(G) for each sample varies from$\sim1580.8$cm$^{-1}$ to $\sim1583.5$cm$^{-1}$, indicating different strains. For example, since Pos(G)$\sim1581.5$cm$^{-1}$ for zero strain\cite{FerrPRL2006,PiscPRL2004}, this implies sample S2 has an average tensile $\epsilon\sim 0.03\%$ while sample S4 has an average compressive $\epsilon\sim 0.09\%$. The maximum absolute strain is $\epsilon \sim 0.1\%$ in sample S4. 

Ref.\citenum{NeumNATC2015} reported a Raman map of SLG encapsulated in hBN containing blisters. Pos(G) and Pos(2D) varied by$\gtrsim$5cm$^{-1}$ and $\gtrsim$15cm$^{-1}$ across $\sim$200$\mu$m$^{2}$. $\Delta\epsilon$ in Ref.\citenum{BansSCIA2015} was$\sim0.2-0.3\%$, around one order of magnitude larger than in our samples. Ref.\citenum{NeumNATC2015} detected FWHM(2D)$>$20cm$^{-1}$ over blisters, as compared to blister-free regions where they found FWHM(2D)$<$20cm$^{-1}$. A similar behavior can be observed in Fig.\ref{fig:Fig4}d, where the blisters in the portion of the sample cleaned at $110^{\circ}$C appear as spots with increased FWHM(2D) in the Raman map, while FWHM(2D) in the portion cleaned at $180^{\circ}$C is homogeneous (spread$\sim$1cm$^{-1}$) and narrow ($<$17cm$^{-1}$).
\begin{figure*}
\centerline{\includegraphics[width=180mm]{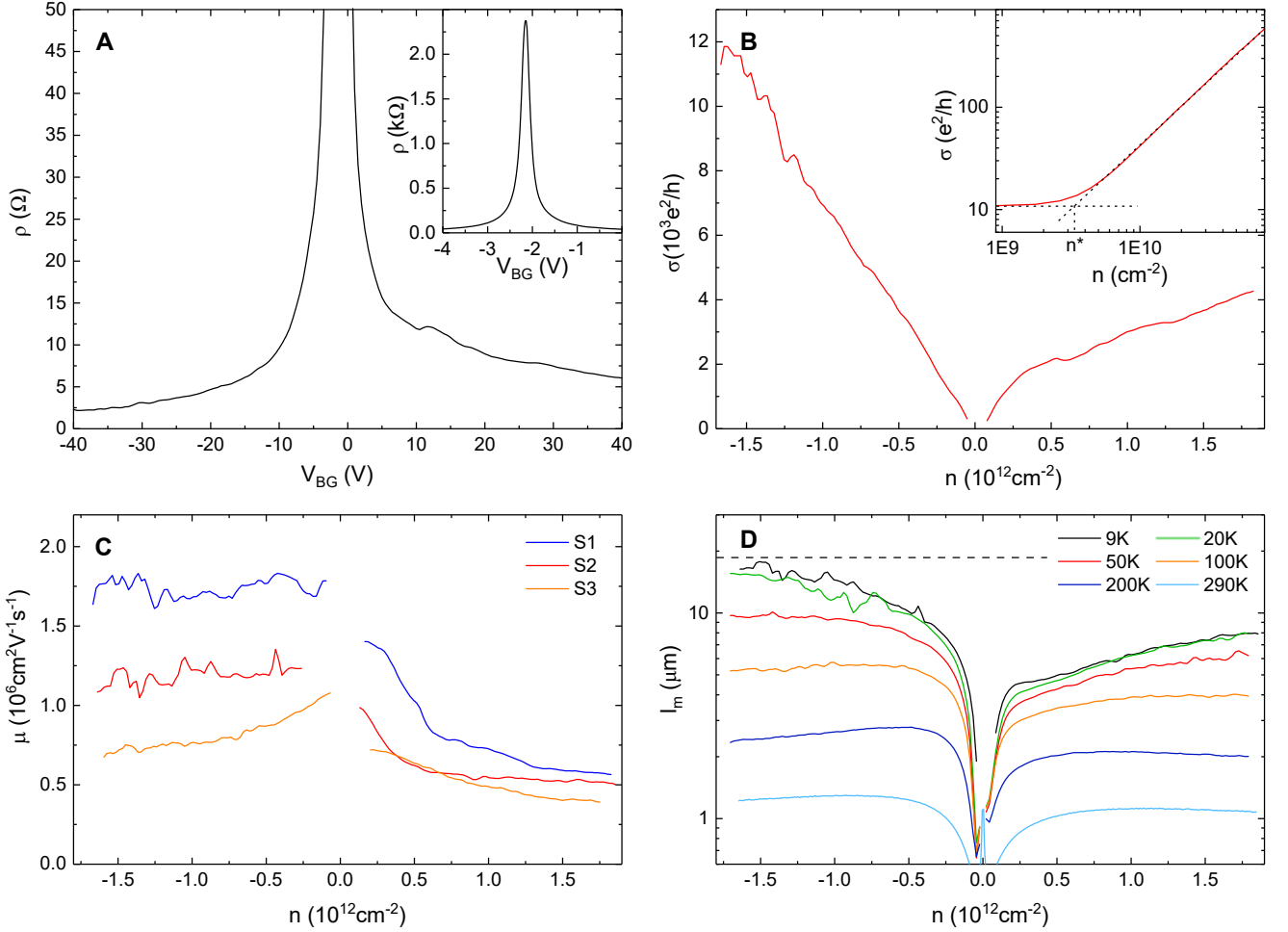}}
\caption{4 terminal measurements of encapsulated SLG at 9K. a) Resistivity vs. gate voltage for sample S1. Inset: Resistivity close to CNP shown on a separate scale for clarity. b) Conductivity vs. n for the same sample. Inset: conductivity vs n close to CNP in logarithmic scale, from which the disorder induced charge inhomogeneity, $n^{*}$, is extracted, as for Ref.\citenum{DuNATN2008,CoutPRX2014}. c) n dependent $\mu$ of S1, as well as for S2 and S3 with widths 18$\mu$m and 7.5$\mu$m respectively. d) $l_{m}$ vs. n as a function of T for S2. The dashed line marks the sample width.}
\label{fig:Fig8}
\end{figure*}
\subsection{Transport}
\begin{figure}
\centerline{\includegraphics[width=90mm]{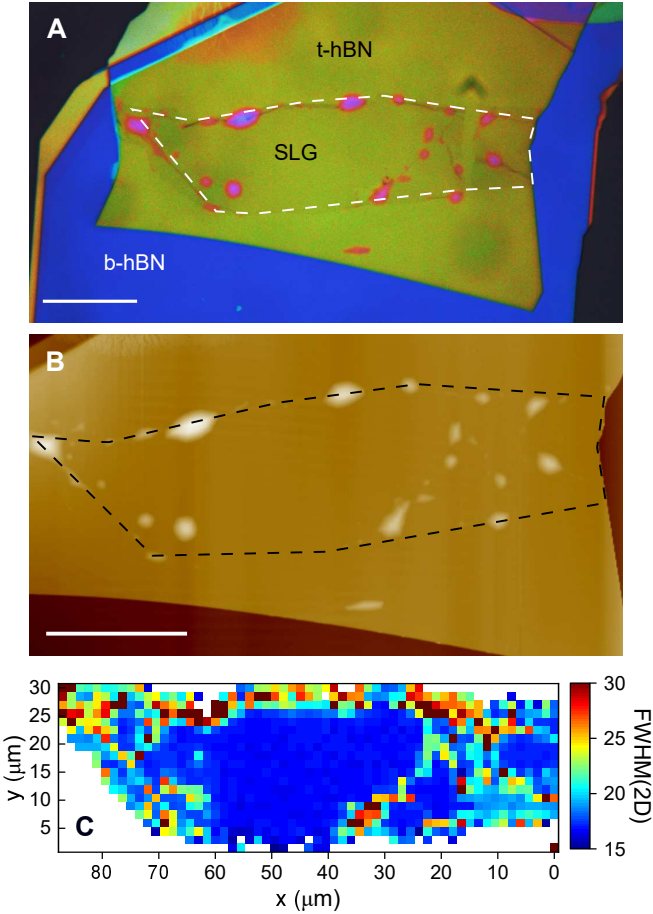}}
\caption{Characterization of LMH produced using SLG exposed to both polymer residuals and solvents. a) False color optical image. SLG is indicated by the white dashed line. Blisters have been pushed to the SLG edges. b) AFM scan of the sample. The black dashed line shows the SLG. c) Spatial map of FWHM(2D) of the sample, taken at an excitation wavelength of 514nm. Scale bars $20\mu$m.}
\label{fig:Fig9}
\end{figure}
\begin{figure*}
\centerline{\includegraphics[width=180mm]{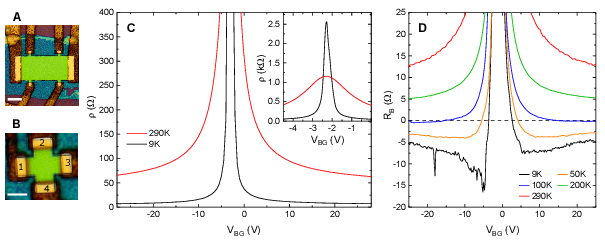}}
\caption{a) Hall bar processed from the sample in Fig.\ref{fig:Fig10}a. The Hall bar corresponds to S7 in Fig. \ref{fig:Fig6}e. Scale bar $5\mu\text{m}$. b)  Hall cross with arm width $2\mu\text{m}$ from the sample in Fig. \ref{fig:Fig9}a. Scale bar $2\mu$m. c) Resistivity at 9 and 290k. Inset: Resistivity at 9 and 290k plotted close to CNP. d) Bend resistance measurements of Hall cross in e as a function of T.}
\label{fig:Fig10}
\end{figure*}
Fig.\ref{fig:Fig7} shows 4 terminal measurements at RT. These are performed using a dual lock-in amplifier (Stanford Research Systems SR810 and SR860), combined with a low noise voltage pre-amplifier (Stanford Research Systems SR860) in a Lakeshore cryogenic probe station at$\sim$3$\times$10$^{-8}$Torr. A bias current of 100nA and a lock-in frequency$\sim$13Hz are used at all T. Fig.\ref{fig:Fig7}a plots the resistivity, $\rho$, as a function of back gate voltage $V_{BG}$. The corresponding conductivity, $\sigma$, as a function of charge carrier density, n, is shown in Fig.\ref{fig:Fig7}b. n as a function of $V_{BG}$ is extracted from a measurement of the Hall voltage with a B=0.5T out of plane magnetic field. From a linear fit of the dependence of n vs. $V_{BG}$ we get a gate capacitance of $C_{ox}=7\times 10^{-5}\text{F}\text{m}^{-2}$. This is in excellent agreement with that calculated assuming a parallel plate capacitor with a bottom hBN flake in series with 285nm SiO$_2$. The bottom hBN thickness is 156nm extracted from AFM. We take its dielectric constant $\epsilon_{r}$=3, considering that values between 2-4 are usually reported\cite{KimACSN2012}. This gives $C_{ox}=7.1\times10^{-5}\text{F}\text{m}^{-2}$. We note that $C_{ox}$ is orders of magnitude smaller than the quantum capacitance of SLG\cite{XiaNATN2009}, which is therefore neglected in the calculations. The sample is highly intrinsic, with a charge neutrality point (CNP) at $V_{BG}$ of $V_{0}=-0.2V$, corresponding to a residual $n_{0}=\left(C_{ox}\slash e\right)V_{0}=9\times10^{9}\text{cm}^{-2}$.

The n dependent $\mu$ is extracted assuming a Drude model of conductivity $\mu=\sigma$/ne, Fig.\ref{fig:Fig7}c. The peak $\mu$ close to the CNP is$\sim180000$cm$^2 V^{-1} \ s^{-1}$, decreasing at higher densities. Of 13 Hall bars with W ranging from 3 up to 24$\mu$m, all exhibit peak RT $\mu>100000$cm$^2 V^{-1} \ s^{-1}$. $\sigma$ of SLG is commonly fit using $\sigma^{-1}=\left(ne\mu_{L} + \sigma_{0}\right)^{-1} + \rho_{s}$, where $\mu_{L}$ represents the contribution from long-range scattering, and $\rho_{s}$ the density independent contribution from short-range scattering\cite{MoroPRL2008,PetrACSN2015,DeanNATN2010,BansSCIA2015}. $\rho_{s}$ results in a sublinear dependence of $\sigma$ with n and therefore decreasing $\mu$ with increasing n. Fitting Fig.\ref{fig:Fig7}b yields $\mu_{L}=217000$cm$^2 V^{-1} \ s^{-1}$ and $\rho_{s}=33\Omega$. For ultra-high $\mu$ encapsulated samples at RT, the dominant contribution to $\rho_{s}$ has been attributed to electron-phonon scattering $\rho_{e\text-ph}$\cite{WangSCI2013}, which sets an upper bound on the achievable $\mu =1\slash$ne$\rho_{e\text-hp}$. At T=290K the theoretically predicted $\rho_{e\text-ph}\sim30-40\Omega$\cite{ParkNL2014,SohiCMMP2014} is consistent with our extracted $\rho_{s}=33\Omega$. For $n=9\times10^{12}\text{cm}^{-2}$ we measure $\mu\sim19000$cm$^2 V^{-1} \ s^{-1}$, consistent with the phonon limit$\sim20000$cm$^2 V^{-1} \ s^{-1}$ calculated for this n\cite{ParkNL2014}.

$\rho$ and $\sigma$ at 9K (corresponding to the base T for our measurement system) for the LMH in Fig.\ref{fig:Fig7} (S1) are plotted in Figs\ref{fig:Fig8}a,b. The n dependent $\mu$ of S1, as well as two others (S2 and S3 with W=18 and 7.5$\mu\text{m}$), are shown in Fig.\ref{fig:Fig8}c. All have a peak $\mu>$1$\times$10$^6$cm$^2 V^{-1} \ s^{-1}$. In S1 $\mu$ reaches 1.8$\times$10$^6$cm$^2 V^{-1} \ s^{-1}$. We note that in S1, for p-doping, $\mu$ remains above 1.5$\times$10$^6$cm$^2 V^{-1} \ s^{-1}$ even at n$>1\times10^{12}\text{cm}^{-2}$. E.g., $\mu$=1.7$\times$10$^6$cm$^2 V^{-1} \ s^{-1}$ at n=1.5$\times10^{12}\text{cm}^{-2}$, in close agreement with ballistic measurements on SLG encapsulated in hBN at similar n\cite{MayoNL2011,WangSCI2013,BansNL2016}. Assuming diffusive transport, i.e. $l_{m}<W$\cite{MayoNL2011}, we can write $l_{m}=\left(h\slash 2e^{2}\right)\sigma\left(1\slash\sqrt{n\pi}\right)$\cite{HwangPRL2007}, meaning $l_{m}\propto \sigma$ for a given n. As the lateral dimensions of the sample constrain $l_{m} \lesssim W$\cite{MayoNL2011,KretNL2014}, W sets an upper bound on the achievable $\sigma$, and therefore $\mu$, for a particular value of n. Achieving $\mu$=1.7$\times$10$^6$cm$^2 V^{-1} \ s^{-1}$ at n=1.5$\times10^{12}\text{cm}^{-2}$ can therefore be seen as a direct result of $W>20\mu\text{m}$. $l_{m}$ of S2 is plotted in Fig.\ref{fig:Fig8}b between 9 and 290K. The sample width is marked by a dashed line, showing that $l_{m}<W$ for all n and T, indicating transport remains diffusive\cite{MayoNL2011}.

The CNP FWHM, $\delta V$, as a function of carrier density, $\delta\text{n}= \left(C_{ox}\slash e\right)\delta V$, places an upper bound on the disorder induced charge inhomogeneity, $n^{*}$\cite{BoloSSC2008,CoutPRX2014,BansSCIA2015}. For the sample in Fig.\ref{fig:Fig8}, $\delta n=10^{10}$cm$^{-2}$, almost an order of magnitude lower than typical reports for SLG on hBN\cite{DeanNATN2010,KretNL2014}. A more precise $n^{*}$ can be extracted by fitting the linear and plateau regions of $\sigma$ at the CNP\cite{DuNATN2008,CoutPRX2014} (inset in Fig.\ref{fig:Fig8}b), giving $n^{*}=3.5\times 10^{9}\text{cm}^{-2}$. $n^{*}$ provides a measure of the spatial inhomogeneity of the carrier density close to the CNP \cite{MartNATP2007}, which arises due to disorder (e.g. local variations in strain\cite{GibeCMMP2012}, or chemical doping\cite{MayoNL2012}). Lower $n^{*}$ are indicative of less disordered, more homogeneous samples. Our $n^{*}=3.5\times 10^{9}\text{cm}^{-2}$ is approximately three times lower than typical reported $n^{*}>1\times10^{10}\text{cm}^{-2}$ for SLG encapsulated in hBN\cite{KretNL2014,BansSCIA2015}.
\subsection{Cleaning of polymer-contaminated samples}
Our method also works for LMHs where the SLG surface is exposed to polymers and solvent before encapsulation, which is a common occurrence when the SLG undergoes lithographic processing\cite{IshiNL2017,PizzNATC2016}, or during a wet and (or) polymer assisted transfers used to process large area SLG films\cite{BansSCIA2015,CaldEST2010,MattJMC2011,GaoNATC2012,WangACSN2011}. To demonstrate this, we spin coat PMMA ($8\%$ in Anisole, 495K molecular weight, spun at 4000rpm for 60s) onto exfoliated SLG on SiO$_{2}$+Si. PMMA is then removed by rinsing in Acetone/IPA. SLG is then encapsulated following the same procedure as in Fig.\ref{fig:Fig1}. The only modification is that cleaning (Fig.\ref{fig:Fig1}g) is performed at 250$^\circ$C, as we find the blisters remain immobile at 180$^\circ$C in these samples. This is in agreement with the analytical model discussed above, which predicts optimal cleaning at $T_{0}\sim250^{\circ}$C. This need for higher T cleaning could be attributed to the increased amount of contaminants trapped at the interfaces in these samples.

Fig.\ref{fig:Fig9}a is an optical image of the cleaned LMH, with the SLG location indicated by a white dashed line. Fig.\ref{fig:Fig9}b is an AFM scan, with the SLG marked by a dashed black line, from which it can be seen that the blisters have been pushed to the SLG edge. A few blisters remain within the SLG, pinned by folds and cracks. A FWHM(2D) map across the sample is in Fig.\ref{fig:Fig9}c. The blister-free region exhibits homogeneous (spread $<1\text{cm}^{-1}$) and narrow ($\sim 17\text{cm}^{-1}$) FWHM(2D), consistent with uncontaminated SLG (see Fig.\ref{fig:Fig4}d).

We measure $\mu$ of our initially polymer contaminated SLG, by processing them into 4-terminal geometries as detailed earlier. Figs.\ref{fig:Fig10}a,b show a Hall bar and Hall cross processed from the sample in Fig.\ref{fig:Fig9}. Fig.\ref{fig:Fig10}c plots $\rho$ extracted from a Hall bar at 290 and 9K. We get $\mu\sim$150000cm$^2 V^{-1} \ s^{-1}$ at 290K and 1.3$\times$10$^{6}$cm$^2 V^{-1} \ s^{-1}$ at 9K, and $n^{*} \sim5.5\times10^{9}\text{cm}^{-2}$. For comparison both Refs.\citenum{DeanNATN2010,PetrNL2012} used SLG on hBN (un-encapsulated) where the SLG surface was also exposed to polymers and solvents, and reported $\mu\sim50000-100000$cm$^{-2}$V$^{-1}$s$^{-1}$ at T$<$10K. Ref.\citenum{KretNL2014} used encapsulated SLG in hBN, where the SLG was exposed to solvent and polymer residue before encapsulation, achieving $\mu\sim150000$cm$^{-2}$V$^{-1}$s$^{-1}$ at T$<$10K. These results both achieve $\mu$ an order of magnitude lower than in our cleaned samples, clearly demonstrating the effectiveness of our technique. 

In order to further confirm the cleanliness of the interfaces in the LMH containing initially polymer contaminated SLG we also investigate ballistic transport. To the best of our knowledge micrometer scale ballistic transport in SLG has only reported in the highest quality SLG encapsulated in hBN samples\cite{MayoNL2011,WangSCI2013,BansNL2016} where the interfaces are clean\cite{WangSCI2013,HaigNATM2012}, and $\mu\gg100000$cm$^{-2}$V$^{-1}$s$^{-1}$. Ballistic transport is commonly probed using bend resistance measurements\cite{TakaSSC1988,BeenPRL1989,MayoNL2011,WangSCI2013,BansNL2016}, where current is applied around a bend in a sample and the corresponding voltage developed measured. We perform these on the Hall cross shown in Fig.\ref{fig:Fig10}b, with arm width $H=2\mu$m. A current is passed from contact 1 to 2 ($I_{1,2}$), while measuring the voltage drop between contacts 4 and 3 ($V_{4,3}$). The bend resistance is defined as $R_{B} = V_{4,3}/I_{1,2}$\cite{MayoNL2011}. For diffusive transport, where $l_{m}\ll H$, carriers travel diffusively around the bend, and $R_{B}$ is positive and determined by the van-der-Pauw formula $R_{B} = \rho\pi\slash \ln2$\cite{MayoNL2011}. However if $l_{m}$ exceeds $H$, carriers injected at contact 1 travel ballistically to 3, resulting in negative $R_{B}$\cite{TakaSSC1988,BeenPRL1989,MayoNL2011}. A negative $R_{B}$ therefore imposes $l_{m}>H$, from which a lower bound on $\mu$ can be calculated from $\mu=(2e\slash h)l_{m}\sqrt{\pi\slash n}$ where $l_{m}>H$\cite{WangSCI2013,BansNL2016}. Fig.\ref{fig:Fig10}d plots $R_{B}$ as a function of T. At 9K and $n=1.1\times10^{12}\text{cm}^{-2}$ we estimate $\mu>$520000cm$^2 V^{-1} \ s^{-1}$. At 290K $\mu$ extracted diffusively yields $\mu\sim$150000cm$^2 V^{-1} \ s^{-1}$. These measurements are consistent with those on the highest $\mu$ encapsulated SLG in literature where RT $\mu\sim 150000\text{cm}^2\text{V}^{-1}\ s^{-1}$\cite{MayoNL2011,WangSCI2013,KretNL2014}, demonstrating that exposure of the SLG surface to polymers or solvents before encapsulation poses no limitations once the appropriate cleaning procedure is used.
\section{Conclusions}
We developed a transfer method that allows blisters to be mechanically manipulated, and removed from LMHs. This enabled us to achieve blister-free regions of SLG encapsulated in hBN limited only by the size of the exfoliated flakes. We achieved mobilities up to$\sim 180000$ cm$^2$ V$^{-1}$s$^{-1}$ at room temperature, and $\sim1.8\times 10^{6}$ cm$^2$V$^{-1}$s$^{-1}$ at 9K. Our method can be used to clean encapsulated samples assembled with polymer contaminated SLG, and these show equivalent mobilities, up to $\sim150000$ cm$^2$V$^{-1}$s$^{-1}$ at room temperature, indicating that the polymer and solvent residuals can be removed from the SLG/hBN interface. Our approach is general and can be used for other LMHs.
\section{Acknowledgements}
We thank Duhee Yoon for useful discussions. We acknowledge funding from EU Graphene Flagship, ERC grant Hetero2D, EPSRC grants EP/L016087/1, EP/K01711X/1, EP/K017144/1, Wolfson College, the Elemental Strategy Initiative conducted by MEXT and JSPS KAKENHI Grant JP15K21722.

\end{document}